\documentclass[aps,prd,twocolumn,superscriptaddress]{revtex4}
\usepackage{epsfig,epsf}
\usepackage{amsmath}
\usepackage{amsthm}
\usepackage{amsfonts}
\usepackage{amssymb}
\usepackage{dsfont}
\usepackage{multirow}
\usepackage{appendix}
\usepackage{slashed}
\usepackage[active]{srcltx}
\usepackage{psfrag}

\begin{document}

\title{ Hidden Bottom Pentaquark States with Spin 3/2 and 5/2}

\date{\today}
\author{K.~Azizi}
\affiliation{Physics Department, Do\u gu\c s University,
Ac{\i}badem-Kad{\i}k\"oy, 34722 Istanbul, Turkey}
\affiliation{School of Physics, Institute for Research in Fundamental Sciences (IPM),
P.~O.~Box 19395-5531, Tehran, Iran}
\author{Y.~Sarac}
\affiliation{Electrical and Electronics Engineering Department,
Atilim University, 06836 Ankara, Turkey}
\author{H.~Sundu}
\affiliation{Department of Physics, Kocaeli University, 41380 Izmit, Turkey}

\begin{abstract}
Theoretical investigations of the pentaquark states which were recently discovered 
provide  important information on their nature and structure. It
is necessary to study the spectroscopic parameters like masses and residues of particles belong to the class of
pentaquarks  and ones having  similar structures. The mass and pole residue are quantities
which emerge as the main input parameters in exploration of  the electromagnetic,
strong and weak interactions of the pentaquarks with other hadrons in many frameworks. This work deals with a QCD
sum rule analysis of the spin-$3/2$ and spin-$5/2$ bottom pentaquarks with   both positive and
negative parities aiming to evaluate their masses and residues. In calculations, the pentaquark
states are modeled by molecular-type interpolating currents: for particles with $J=5/2$ a
mixing current is used. We compare the results obtained in this work with the existing predictions of
other theoretical studies. The predictions on the masses may shed light on experimental searches of the bottom pentaquarks. 

\end{abstract}

\maketitle

\section{Introduction}

The announcement  by the LHCb Collaboration~\cite{Aaij:2015tga}
on the observation of the two charmed pentaquark states  placed the subject under the spotlight
in both theoretical and experimental sides. The non-conventional internal quark structure of these states, which are  excluded  neither by  the naive quark model  nor by QCD, puts them at the focus of increasing interests. Many experimental studies have conducted to prove existence of these particles as well as to explore their internal structures. Parallel theoretical studies on the nature of these exotic baryons are in progress. 

The experimental searches for the pentaquark states have a long and controversial story. We refrain from
listing all those searches  and refer the reader to Ref.~\cite{Azizi:2016dhy} and references therein for a full history.
Although their existence was predicted many decades ago by  Jaffe~\cite{Jaffe:1976ig} and their properties were worked out in many theoretical studies (see for instance Refs.~\cite{Gignoux:1987cn,Hogaasen:1978jw,Strottman:1979qu,Lipkin:1987sk,
Fleck:1989ff,Oh:1994np,Chow,Shmatikov,Genovese,Lipkin2,Lichtenberg}), the searches on the pentaquarks ended up in positive results recently and the two pentaquark states, $P_c^+(4380)$ and $P_c^+(4450)$, were reported by LHCb Collaboration in 2015 in the
$\Lambda_b^0\rightarrow J/\psi K^-p$ decays with masses $4380\pm 8\pm 29$ MeV and $4449.8\pm 1.7\pm 2.5$ MeV, spins $3/2$ and $5/2$ and decay widths $205\pm 18\pm
86$ MeV and $39\pm 5\pm 19$ MeV, respectively~\cite{Aaij:2015tga}. There are other states which are interpreted as other possible pentaquark states. In Refs.~\cite{Kim:2017jpx} some of the newly observed $\Omega_c$ states  by LHCb~\cite{Aaij:2017nav}  were considered among  possible pentaquark states. Also, in Ref.~\cite{He:2017aps} the states $N(1875)$ and $N(2100)$ were stated to be possible strange partners of $P_c^+(4380)$ and $P_c^+(4450)$, respectively.

The observation of LHCb  boosted  intense theoretical works to provide an explanation of the properties of these states. Via different models such as, diquark-triquark
model~\cite{Wang:2016dzu,Zhu:2015bba,Lebed:2015tna},
diquark-diquark-antiquark
model~\cite{Wang:2016dzu,Anisovich:2015cia,Maiani:2015vwa,Ghosh:2015ksa,Wang:2015ava,Wang:2015epa,Wang:2015ixb}, meson baryon molecular
model~\cite{Wang:2016dzu,Roca:2015dva,Chen:2015loa,Huang:2015uda,Meissner:2015mza,Xiao:2015fia,
He:2015cea,Chen:2015moa,Chen:2016heh,Yamaguchi:2016ote,He:2016pfa},  and
topological soliton model~\cite{Scoccola:2015nia}, their properties and substructures were investigated. 
A review on  the multiquark states including pentaquarks and their  possible experimental measurements can be found in  Ref.~\cite{Chen:2016qju} . Some of the recent investigations have considered  other possible substructures for the pentaquark states.
Beside the mass of the hidden-charmed molecular pentaquark states the mass  of  charmed-strange molecular pentaquark states, and  other hidden-charmed molecular pentaquark states, which are named as  $P_c^{'}(4520)$, $P_c^{'}(4460)$, $P_{cs}(3340)$ and $P_{cs}(3400)$,  were  predicted in  Ref.~\cite{Chen:2016heh}.
The same work also contains the predictions on the masses of hidden-bottom pentaquark states with molecular structure. In Ref. \cite{Shimizu:2016rrd} besides the $P_c^+(4380)$ state the possible existence of hidden bottom pentaquarks with a mass around $11080-11110$ MeV and quantum numbers $J^{P}={3/2}^-$ was emphasized,  and it was indicated that there may exist some loosely-bound molecular-type  pentaquarks   with heavy quark contents $c\bar{b}$, $b\bar{c}$ or $b\bar{b}$. For such type of pentaquark states the mass predictions were presented  in Ref. \cite{Liu:2017xzo}. In this work, using a variant of D4-D8 brane model \cite{Sakai:2004cn}, the mass of charmed and bottom pentaquarks  were predicted as $M_{\bar{c}c}=4678$ MeV, $M_{\bar{c}b}=M_{\bar{b}c}=8087$ MeV and $M_{\bar{b}b}=11496$ MeV. See also  references \cite{yeni1,yeni2} for more information on the properties of the charmed and bottom pentaquark states using the coupled-channel unitary approach as well as \cite{yeni3,yeni4,yeni5,yeni6} on the structure of the pentaquarks and triangle singularities. 

In the light  of all  these developments, it is necessary to explore the pentaquarks to gain constructive information on their nature and substructures. If one considers the historical development of the particle physics, the observations of the particles are sequential. The observation of baryons containing a $c$ quark was followed by the observation of similar baryons containing a $b$ quark. Therefore it is natural to expect a possible subsequent observation of the bottom analogues  of the observed pentaquark states. Investigations of their spectroscopic and  electromagnetic properties, as well as their strong and weak decays supply beneficial information for the future experimental searches. In addition to this, further theoretical studies are  helpful to get insights into the nature of these particles, as well as into the dynamics of their strong interactions by comparing the results  with the existing theoretical predictions and experimental data. Starting from this motivation, in this work we extend our previous study on the properties of charmed pentaquarks \cite{Azizi:2016dhy} and  calculate the masses and residues of the pentaquark states $P_b$  with $J=3/2$ and $J=5/2$ by considering both the positive and negative parity states. For this purpose, we use the  QCD sum rule method \cite{Shifman:1978bx,Shifman:1978by},  interpolating currents of the molecular form for the states with $J=3/2$ and a mixed molecular current for those states having $J=5/2$. For the latter we optimize the mixing angle according to the standard prescriptions.

The present work  is organized in the following way. In Sec. II calculations of the mass and residue of hidden bottom pentaquark states are presented.   Section III is devoted to the numerical analysis  and discussion on the obtained results. In Sec. IV we summarize our results and briefly discuss prospects to study decays of the pentaquark states. Some spectral densities used in calculations are moved to the appendix.

\section{The hidden-bottom pentaquark states with $ J= \frac{3}{2}$ and $ J= \frac{5}{2}$ }


This section presents the calculations of the masses and residues of the hidden bottom pentaquark states with
$J=3/2 $ and $ J=5/2 $. In both cases we consider the positive  and negative  parity states.  To begin the calculations, for the state with $ J=3/2 $
we use the following two point correlation function:
\begin{equation}
\Pi _{\mu \nu }(p)=i\int d^{4}xe^{ip\cdot x}\langle 0|\mathcal{T}\{J_{\mu
}^{\bar{B}^*\Sigma_{b}}(x)\bar{J}_{\nu }^{\bar{B}^*\Sigma_{b}}(0)\}|0\rangle ,  \label{eq:CorrF1Pc}
\end{equation}%
where $J_{\mu}^{\bar{B}^*\Sigma_{b}}(x)$ is the interpolating current having the quantum numbers $ J= \frac{3}{2}^{-} $~\cite{Chen:2015moa}. This current couples to both the negative and positive parity particles, and its explicit expression is given as:
\begin{equation}
J_{\mu}^{\bar{B}^*\Sigma_{b}}=[\bar{b}_{d}\gamma_{\mu}d_{d}][\epsilon_{abc}(u_{a}^{T}C\gamma_{\theta}u_{b})\gamma^{\theta}\gamma_{5}b_{c}].
 \label{eq:JJPc}
\end{equation}

For the states with $ J=5/2$ the correlation function has the following form:
\begin{equation}
\Pi _{\mu \nu \rho\sigma}(p)=i\int d^{4}xe^{ip\cdot x}\langle 0|\mathcal{T}\{J_{\mu\nu
}(x)\bar{J}_{\rho\sigma }(0)\}|0\rangle ,  \label{eq:CorrF1}
\end{equation}%
where $J_{\mu \nu}(x)$ is the interpolating current, which also couples to both the positive and negative parity states. This current is chosen as a mixed current composed of $J_{\mu\nu}^{\bar{B}\Sigma_{b}^{*}}$ and $J_{\mu\nu}^{\bar{B}^{*}\Lambda_{b}}$   \cite{Chen:2015moa}:
\begin{eqnarray}
J_{\mu\nu}(x)=\mathrm{\sin\theta} \times J_{\mu\nu}^{\bar{B}\Sigma_{b}^{*}}+\mathrm{\cos\theta} \times J_{\mu\nu}^{\bar{B}^{*}\Lambda_{b}},
\label{eq:CDiq}
\end{eqnarray}
where $\theta$ is a mixing angle that should be fixed, and
\begin{eqnarray}
J_{\mu\nu}^{\bar{B}\Sigma_{b}^{*}}&=&[\bar{b}_{d}\gamma_{\mu}\gamma_{5}d_{d}][\epsilon_{abc}(u_{a}^{T}C\gamma_{\nu}u_{b})b_{c}]+\lbrace\mu\leftrightarrow\nu\rbrace,\nonumber \\
J_{\mu\nu}^{\bar{B}^{*}\Lambda_{b}}&=&[\bar{b}_{d}\gamma_{\mu}u_{d}][\epsilon_{abc}(u_{a}^{T}C\gamma_{\nu}\gamma_{5}d_{b})b_{c}]+\lbrace\mu\leftrightarrow\nu\rbrace.\nonumber \\
 \label{eq:JJ}
\end{eqnarray}%

The  above correlation functions  are calculated in two different ways. On the side of phenomenology, one inserts a complete set of hadronic states with the same quantum numbers as the interpolating currents into the correlation functions. This calculation comes up with results containing hadronic degrees of freedom such as masses and residues. On  QCD side, the same correlation functions are calculated in terms of QCD degrees of freedom. Finally, the coefficients of the same Lorentz structures obtained in both sides are matched and QCD sum rules for the desired physical parameters are obtained.

In the case of states with $J=3/2$, the procedure summarized above  for the physical side leads to the result
\begin{eqnarray}
\Pi_{ \mu \nu }^{\mathrm{Phys}}(p)&=&\frac{\langle 0|J_{\mu }|{\frac{3}{2}}^{+}(p)\rangle
\langle {\frac{3}{2}}^{+}(p)|\bar{J}_{\nu }|0\rangle }{m_{\frac{3}{2}^+}^{2}-p^{2}} \nonumber \\&+&\frac{\langle 0|J_{\mu }|{\frac{3}{2}}^{-}(p)\rangle
\langle {\frac{3}{2}}^{-}(p)|\bar{J}_{\nu }|0\rangle }{m_{\frac{3}{2}^-}^{2}-p^{2}}+\cdots,
\label{eq:physPc}\end{eqnarray}
where $m_{\frac{3}{2}^{+}}$ and $m_{\frac{3}{2}^{-}}$ are the masses of the positive and negative parity particles, respectively. The contributions of the higher states and continuum are represented by the ellipsis in the last equation. The matrix elements in Eq.~(\ref{eq:physPc}) are given in terms of the residues $ \lambda_{\frac{3}{2}^+}$ and  $ \lambda_{\frac{3}{2}^-} $, and corresponding spinors as
\begin{eqnarray}
\langle 0|J_{\mu }|\frac{3}{2}^+(p)\rangle &=&\lambda_{\frac{3}{2}^{+}} \gamma_5 u_{\mu}(p),\nonumber\\
\langle 0|J_{\mu }|\frac{3}{2}^{-}(p)\rangle &=&\lambda_{\frac{3}{2}^{-}} u_{\mu}(p).
\label{eq:ResPc}
\end{eqnarray}
 A similar result for the correlation function corresponding to $J=5/2$ states is obtained:
  \begin{eqnarray}
\Pi _{\mu \nu\rho\sigma }^{\mathrm{Phys}}(p)&=&\frac{\langle 0|J_{\mu \nu}|\frac{5}{2}^{+}(p)\rangle
\langle \frac{5}{2}^{+}(p)|\bar{J}_{\rho\sigma }|0\rangle }{m_{\frac{5}{2}^{+}}^{2}-p^{2}}\nonumber\\&+&\frac{\langle 0|J_{\mu \nu}|\frac{5}{2}^{-}(p)\rangle
\langle \frac{5}{2}^{-}(p)|\bar{J}_{\rho\sigma }|0\rangle }{m_{\frac{5}{2}^{-}}^{2}-p^{2}}\nonumber\\&+& \cdots,
\label{eq:phys}\end{eqnarray}%
with the matrix elements defined as
\begin{eqnarray}
\langle 0|J_{\mu\nu }|\frac{5}{2}^{+}(p)\rangle &=&\lambda_{\frac{5}{2}^{+}}u_{\mu\nu}(p),\nonumber\\\langle 0|J_{\mu\nu }|\frac{5}{2}^{-}(p)\rangle &=&\lambda_{\frac{5}{2}^{-}}\gamma_5u_{\mu\nu}(p).
\label{eq:Res}
\end{eqnarray}
In these equations $m_{\frac{5}{2}^{+}}$ and $m_{\frac{5}{2}^{-}}$ are the masses of the spin-$\frac{5}{2}$ states having positive and negative parities, respectively. Using the  matrix elements parameterized in terms of the masses and residues and  performing the Borel transformation, the physical side is found as
\begin{eqnarray}
&&\mathcal{B}_{p^{2}}\Pi_{\mu \nu  }^{\mathrm{Phys}}(p)=-\lambda_{\frac{3}{2}^{+}}^{2}e^{-\frac{m_{\frac{3}{2}^{+}}^2}{M^2}}(-\gamma_5)({\slashed p}+m_{\frac{3}{2}^{+}})
  g_{\mu\nu}\gamma_5\nonumber \\
&&-
\lambda_{\frac{3}{2}^{-}}^{2}e^{-\frac{m_{\frac{3}{2}^{-}}^2}{M^2}}({\slashed p}+m_{\frac{3}{2}^{-}})
  g_{\mu\nu} +\cdots,
\nonumber \\  \label{eq:CorBorPc}
\end{eqnarray}
for pentaquark states with spin-3/2, with $ M^2 $ being the Borel parameter. Here, $ g_{\mu\nu} $ and $ {\slashed p} g_{\mu\nu}$ are  structures that give contributions to only the spin-3/2 particles. By choosing these structures we eliminate the unwanted spin-1/2 pollution.  In the case of hidden bottom pentaquarks with spin-5/2 we find
\begin{eqnarray}
&&\mathcal{B}_{p^{2}}\Pi_{\mu \nu \rho\sigma}^{\mathrm{Phys}}(p)=\nonumber\\&&\lambda_{\frac{5}{2}^{+}}^{2}
e^{-\frac{m_{\frac{5}{2}^{+}}^2}{M^2}}
(\slashed p+m_{\frac{5}{2}^{+}})
(\frac{g_{\mu\rho}g_{\nu\sigma}+g_{\mu\sigma}g_{\nu\rho}}{2}) \nonumber\\
&&+\lambda_{\frac{5}{2}^{-}}^{2}
e^{-\frac{m_{\frac{5}{2}^{-}}^2}{M^2}}
(\slashed p-m_{\frac{5}{2}^{-}})
(\frac{g_{\mu\rho}g_{\nu\sigma}+g_{\mu\sigma}g_{\nu\rho}}{2})\nonumber\\&&+\cdots \label{eq:CorBor},
\end{eqnarray}
where we kept again only the structures that give contributions to the spin-5/2 particles  and ignored other structures giving contributions to the spin-3/2 and spin-1/2 particles.  

The calculation of the correlation function in terms of QCD degrees of freedom is the next stage of the calculations. In this part, the interpolating currents of the interested states are substituted into the correlation functions  and the quark fields are contracted through the Wick's theorem. This procedures end up in finding the correlation functions  in terms of the light and heavy quark propagators. Using the quark propagators in coordinate space as presented in \cite{Azizi:2016dhy} we apply the Fourier transformation to transfer the calculations to the momentum space. To suppress the contributions of the higher states and continuum we apply the Borel transformation as well as continuum subtraction and use the dispersion integral representation. At the end of  this procedure we obtain the spectral densities as the imaginary parts of the functions 
corresponding to all selected structures.

The calculations of physical and theoretical sides are followed by the selection of the coefficients of the same structures from both sides and their matching to obtain the relevant QCD sum rules that will give us the physical quantities of interest. The final forms of the sum rules are obtained as
\begin{eqnarray}
&&m_{J^+}\lambda_{J^+}^{2}e^{-m_{J^+}^{2}/M^{2}}- m_{J^-}\lambda_{J^-}^{2}e^{-m_{J^-}^{2}/M^{2}}=%
\Pi^{m}_{J},\nonumber\\
&&\lambda_{J^+}^{2}e^{-m_{J^+}^{2}/M^{2}}+\lambda_{J^-}^{2}e^{-m_{J^-}^{2}/M^{2}}=%
s_{J}\Pi^{p}_{J},
\label{eq:srcoupling1}
\end{eqnarray}
where $ J=3/2$ or $5/2 $. In the last equation $ s_{J} $ equals to $ -1 $ for  $ 3/2 $ and $ 1 $  for  $ 5/2 $ states.
The functions $ \Pi^{m}_{3/2} $, $ \Pi^{p}_{3/2} $, $ \Pi^{m}_{5/2} $ and $ \Pi^{p}_{5/2} $ are coefficients of the structures  $ g_{\mu\nu} $, $ {\slashed p} g_{\mu\nu}$, $  (g_{\mu\rho}g_{\nu\sigma}+g_{\mu\sigma}g_{\nu\rho})/2 $ and $ {\slashed p} (g_{\mu\rho}g_{\nu\sigma}+g_{\mu\sigma}g_{\nu\rho})/2 $, respectively on the side of QCD. These functions  are  written in terms of the spectral densities as: 
\begin{equation}
\Pi^{j}_{J}=\int^{s_0}_{4m^2_b}ds\rho^{j}_{J}(s)e^{-s/M2},
\label{spectral}
\end{equation}%
where,  $ j=m$ or $p $. The spectral densities can also be written in terms of the perturbative and nonperturbative parts as
\begin{equation}
\rho^{j}_{J} (s)=\rho_{J} ^{\mathrm{j,pert.}}(s)+\sum_{k=3}^{6}\rho^{j}_{J,k}(s),
\label{eq:A1}
\end{equation}%
where  $\rho^{j}_{J,k}(s)$ represents the   nonperturbative contributions to the spectral densities. As examples, we present the perturbative and nonperturbative parts of the spectral densities corresponding to the structures $g_{\mu\nu}$  and  $  (g_{\mu\rho}g_{\nu\sigma}+g_{\mu\sigma}g_{\nu\rho})/2 $ in terms of the the integrals over the Feynman parameters $x$ and $ y $ in the appendix.

As is seen, the sum rules contain four unknowns in each case which include the masses and residues of considered states. We need two extra equations in each case that are obtained by  applying a derivative with respect to  $\frac{1}{M^2}$ to both sides of the above equations. By simultaneous solving of the obtained equations' sets one can obtain the masses and residues of the particles with both parities in terms of the QCD degrees of freedom as well  the Borel parameter, continuum threshold and mixing angle in the case of spin-5/2 particles.

\section{Numerical results}
\begin{table}[tbp]
\begin{tabular}{|c|c|}
\hline\hline
Parameters & Values \\ \hline\hline
$m_{b}$ & $(4.78\pm 0.06)~\mathrm{GeV}$ \\
$\langle \bar{q}q \rangle $ & $(-0.24\pm 0.01)^3$ $\mathrm{GeV}^3$  \\
$m_{0}^2 $ & $(0.8\pm0.1)$ $\mathrm{GeV}^2$ \\
$\langle \overline{q}g_s\sigma Gq\rangle$ & $m_{0}^2\langle \bar{q}q \rangle$
\\
$\langle\frac{\alpha_sG^2}{\pi}\rangle $ & $(0.012\pm0.004)$ $~\mathrm{GeV}%
^4 $\\
\hline\hline
\end{tabular}%
\caption{Some input parameters used in the calculations.}
\label{tab:Param}
\end{table}
The input  parameters that are needed in the  numerical analyses of the obtained  sum rules in the previous section are collected in table I. Note that in the numerical calculations we use the $ b $ quark pole mass and the masses of light $u$ and $d$ quarks are taken as zero. 
 It is well known that the parameters of the bottom systems depend  on the $ b $ quark mass, considerably. However, our numerical analyses show that the results of physical quantities under consideration show more stability with respect to the changes of the auxiliary parameters when the $ b $ quark pole mass is used compared to the one that the $  b$  quark running mass in $ \overline{\text{MS}} $ scheme is taken into account.  Our analyses also show that when we use  the $ b $ quark pole mass we achieve higher pole contributions in all channels compared to the case of  $ b $ quark running mass. Therefore, we choose the $ b $ quark pole mass to numerically analyze the obtained sum rules.
 
 The next step is to determine the working intervals for two auxiliary parameters, namely the continuum  threshold $s_0$ and the Borel parameter $M^{2}$. For the determination of the Borel window, the convergence of the series of OPE and the adequate suppression of the contributions coming from the higher states and continuum are taken into account. These lead to the interval
\begin{equation}
11\ \mathrm{GeV}^{2}\leq M^{2}\leq 16\ \mathrm{GeV}^{2}.
\end{equation}
for both states. The pole dominance and OPE convergence are also considered in  determination of the working region for the threshold parameter which is obtained as
\begin{equation}
141\,\,\mathrm{GeV}^{2}\leq s_{0}\leq 145\,\,\mathrm{GeV}^{2},
\end{equation}
for $J=\frac{3}{2}$ states with both  parities and
\begin{equation}
142\,\,\mathrm{GeV}^{2}\leq s_{0}\leq 146\,\,\mathrm{GeV}^{2},
\end{equation}
for $J=\frac{5}{2}$ states with negative and positive parities. Note that the above intervals for the continuum threshold are valid for both the  $ b $ quark pole mass and  running mass in $ \overline{\text{MS}} $ scheme.
The calculation of the desired parameters of spin-5/2 states with the chosen interpolating current also requires determination of another auxiliary parameter which is the mixing angle entering the interpolating current. We look for a working interval for this parameter such that our results depend on it relatively weakly. Our analyses show that the dependence of the results on $\cos\theta$ in the region $ -0.5 \leqslant \cos\theta\leqslant 0.5 $ for both the masses of the positive and negative parity pentaquarks with $J =\frac{5}{2} $ is  weak (see figure 1). We use $\cos\theta $ to easily sweep the whole region by varying it in the interval $ [-1,1] $. It is worth nothing that the pole quark mass together with the above intervals for the auxiliary parameters lead to maximally 78\% and 79\% pole contributions in spin-3/2 and spin-5/2 channels, respectively, which nicely satisfy the requirements of the QCD sum rules calculations. 
\begin{widetext}

\begin{figure}[h!]
\begin{center}
\includegraphics[totalheight=5cm,width=7cm]{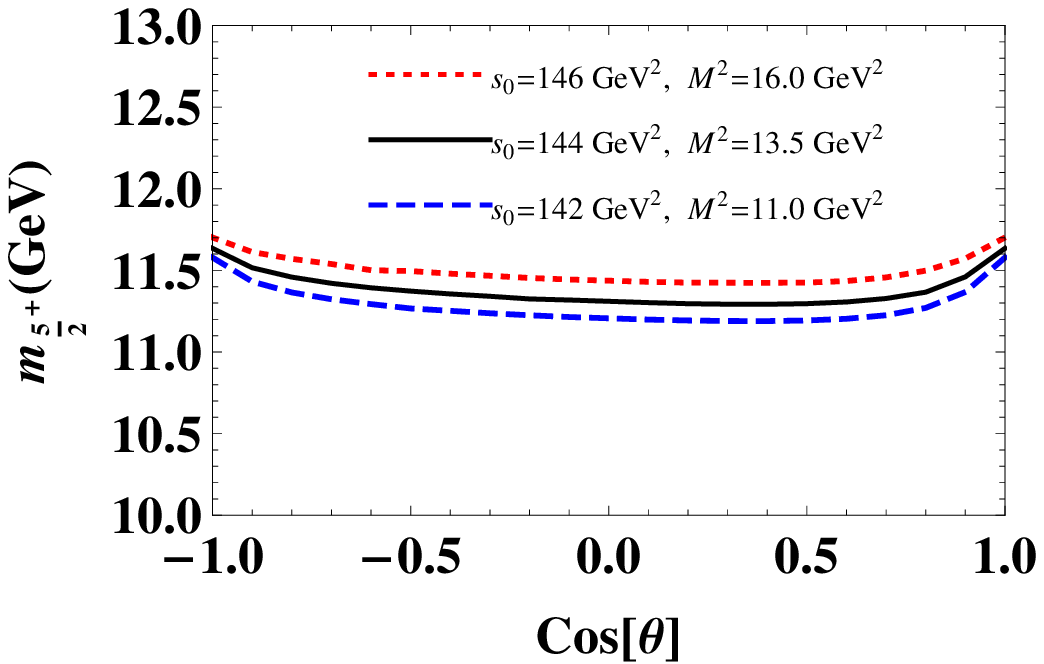}
\includegraphics[totalheight=5cm,width=7cm]{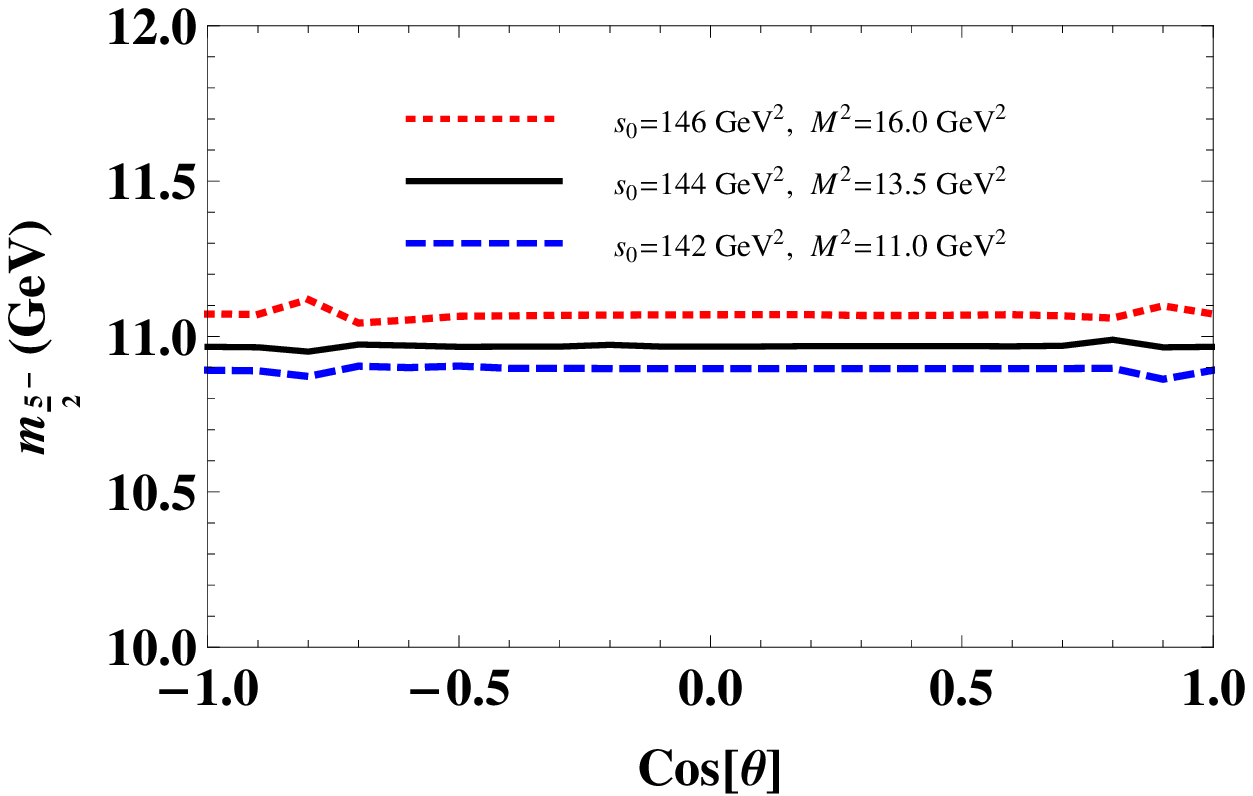}
\end{center}
\caption{\textbf{Left:} The mass  of the  pentaquark with $J^{P} =\frac{5}{2} ^{+} $  as a function of $\cos\theta$  at different fixed values of the continuum threshold and Borel parameter.
 \textbf{Right:}
 The mass  of the  pentaquark with $J^{P} =\frac{5}{2} ^{-} $  as a function of $\cos\theta$  at different fixed values of the continuum threshold and Borel parameter. } \label{masscos52theta}
\end{figure}

\end{widetext}

As examples, the dependence of the  masses and residues of the hidden bottom pentaquark states with spin-5/2  on  $M^2$ at different fixed values of  $s_0$  are shown in Figs.~\ref{mass52Msq} and \ref{residue52Msq}. From these figures it can be seen that the choices for the working intervals ensure the requirement of weak dependency of the results on these auxiliary parameters.

In this part, to see how the results depend on the $ b $ quark mass, as an  example,  we compare the mass of the pentaquark state with $J^{P} =\frac{3}{2} ^{+} $ obtained via $ b $ quark pole mass (left panel) and $  b$  quark running mass in $ \overline{\text{MS}} $ scheme (right panel) as a function of  $s_0$  at different fixed values  of $M^2$ in Fig.~\ref{comparision}. From this figure it is obvious that the mass of this state changes with amount of 3.2\% in average when switching from the pole mass to the running mass. This amount becomes considerably large in the case of residues.  However, as is seen from this figure, the mass of this state is more stable with respect to the changes of the auxiliary parameters when the $ b $ quark pole mass is used compared to the case of $  b$  quark running mass. The  masses  of  other states and especially the residues of all particles under consideration are also found to be more stable for the case of  $ b $ quark pole mass.   

\begin{widetext}

\begin{figure}[h!]
\begin{center}
\includegraphics[totalheight=5cm,width=7cm]{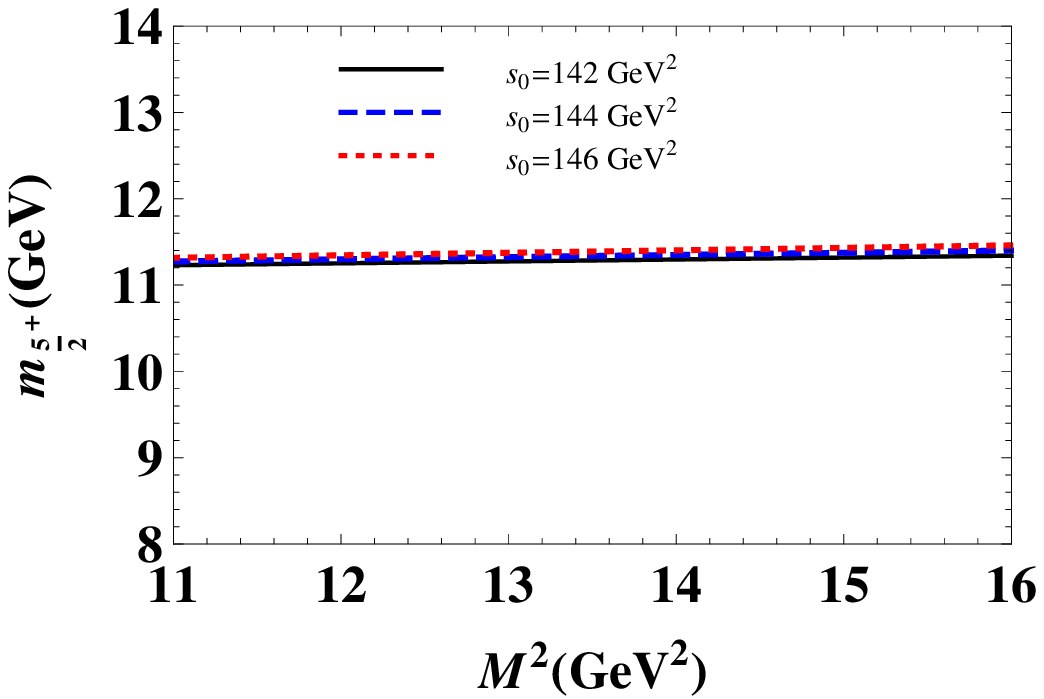}
\includegraphics[totalheight=5cm,width=7cm]{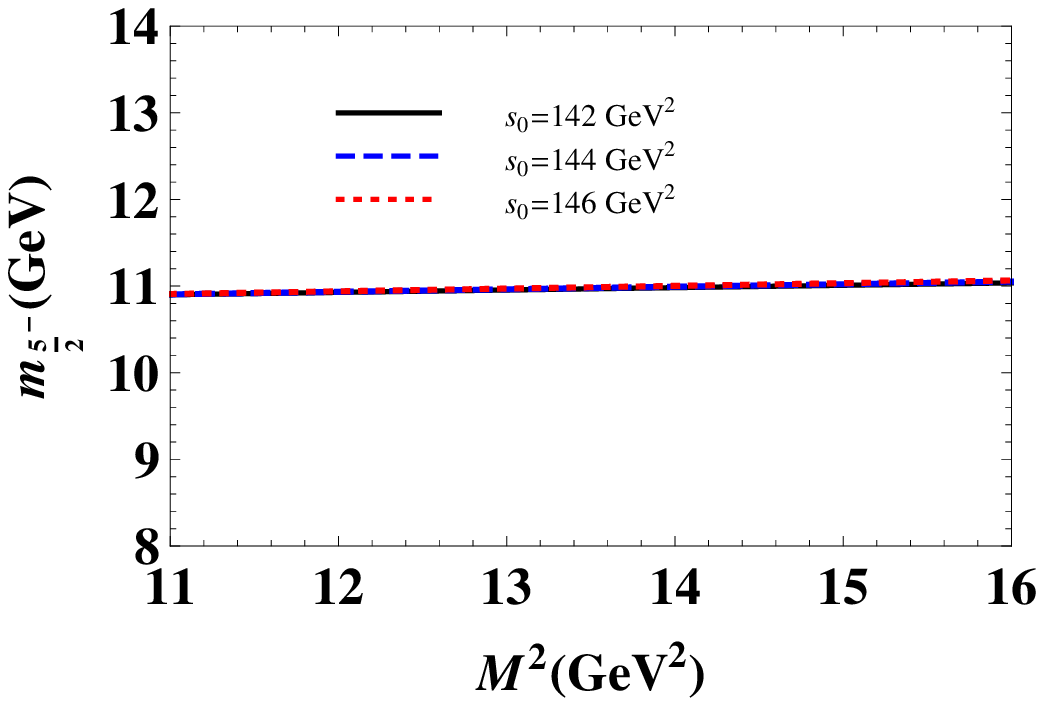}
\end{center}
\caption{\textbf{Left:} The mass  of the  pentaquark with $J^{P} =\frac{5}{2} ^{+} $  as a function of Borel
parameter $M^2$  at different fixed values of the continuum threshold. \textbf{Right:}
 The mass  of the  pentaquark with $J^{P} =\frac{5}{2} ^{-} $  as a function of Borel
parameter $M^2$  at different fixed values of the continuum threshold. } \label{mass52Msq}
\end{figure}
\begin{figure}[h!]
\begin{center}
\includegraphics[totalheight=5cm,width=7cm]{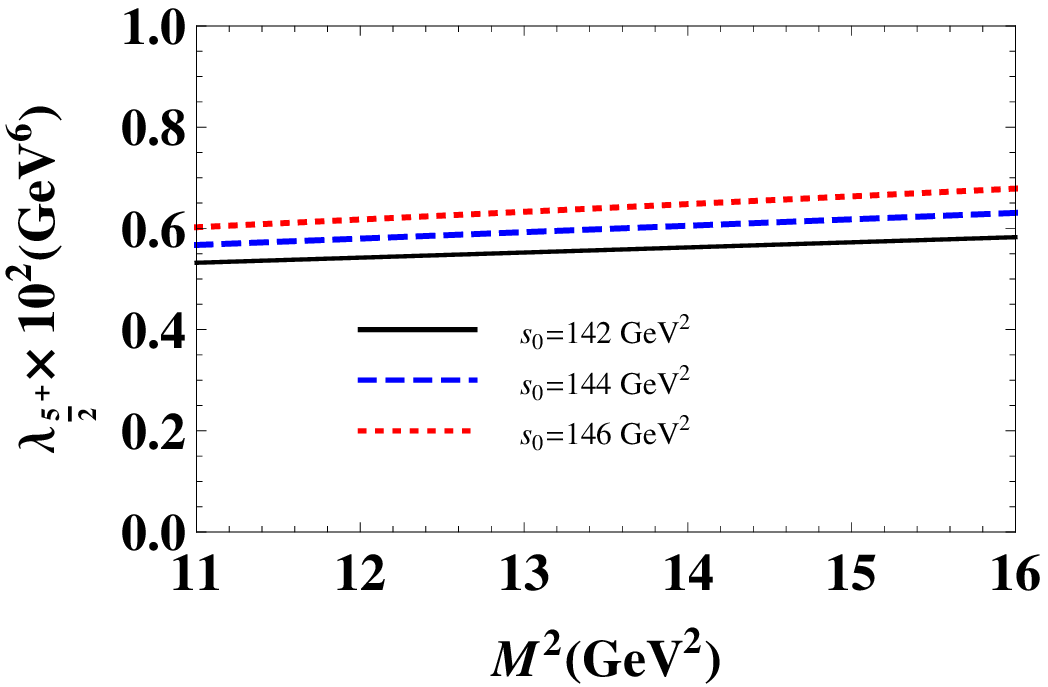}
\includegraphics[totalheight=5cm,width=7cm]{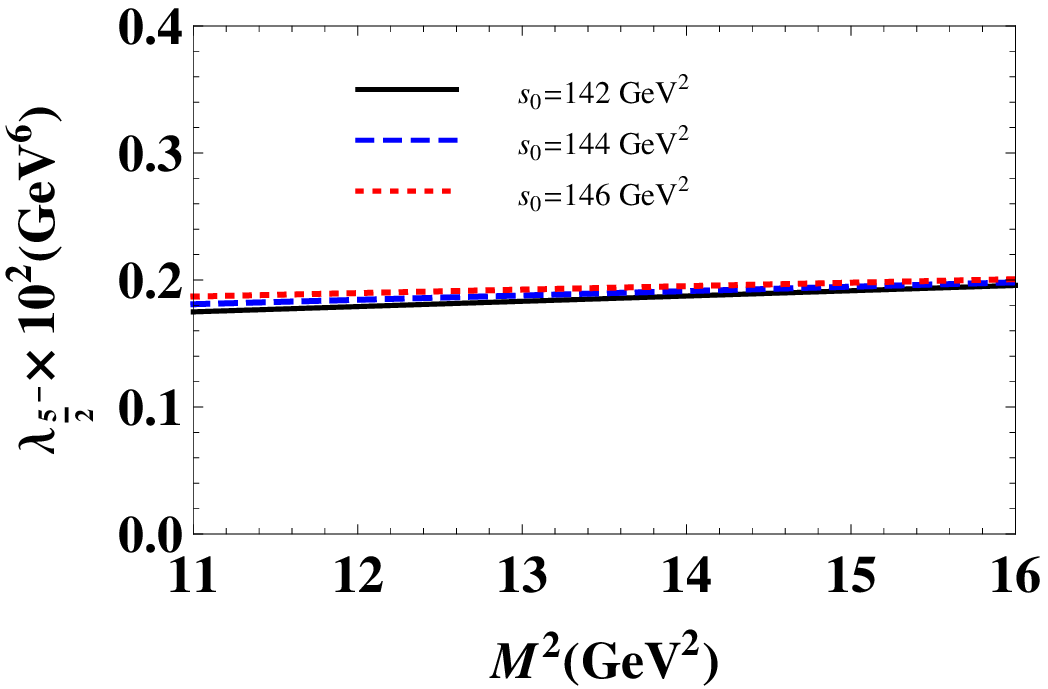}
\end{center}
\caption{\textbf{Left:} The residue of the pentaquark with $J^{P} =\frac{5}{2} ^{+} $  as a function of  $M^2$  at different fixed values  of the continuum threshold.
 \textbf{Right:}
 The residue of the pentaquark with $J^{P} =\frac{5}{2}  ^{-}$  as a function of  $M^2$  at different fixed values of the continuum threshold.} \label{residue52Msq}
\end{figure}
\begin{figure}[h!]
\begin{center}
\includegraphics[totalheight=5cm,width=7cm]{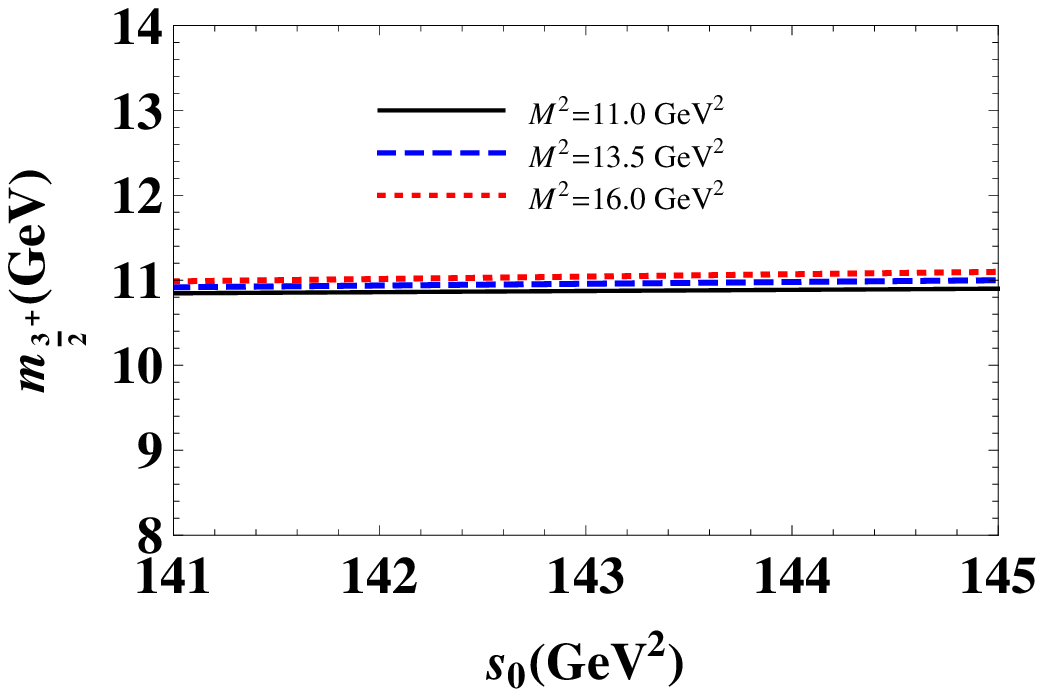}
\includegraphics[totalheight=5cm,width=7cm]{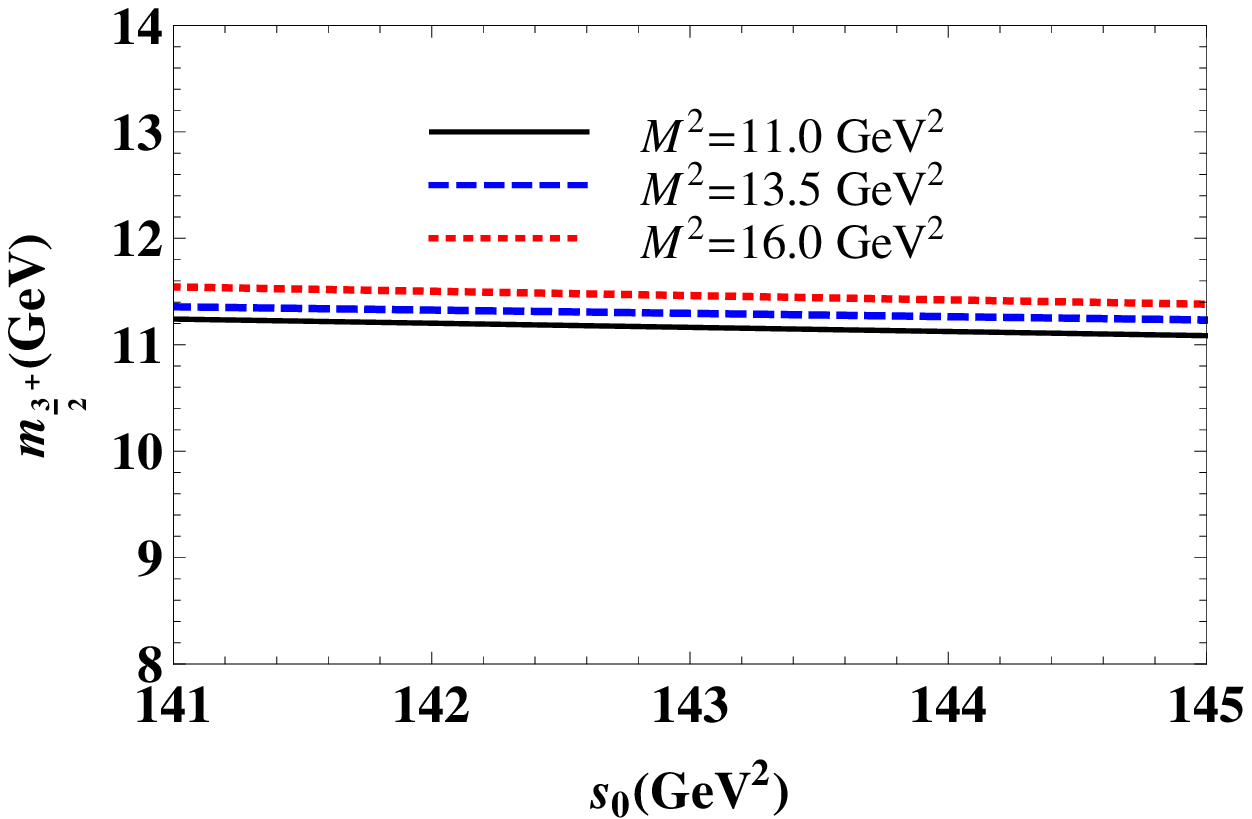}
\end{center}
\caption{\textbf{Left:} The mass of the pentaquark with $J^{P} =\frac{3}{2} ^{+} $  as a function of  $s_0$  at different fixed values  of $M^2$ using the $ b $ quark pole mass.
 \textbf{Right:}
 The mass of the pentaquark with $J^{P} =\frac{3}{2}  ^{+}$  as a function of  $s_0$  at different fixed values of $M^2$ using the $ b $ quark running mass.} \label{comparision}
\end{figure}

\end{widetext}
%
%
Having established the intervals required for the auxiliary parameters $M^2$ and $s_0$, in the next step these regions are applied to evaluate the  masses, $m_{P_b}$ and residues, $\lambda_{P_b}$ of the  states under consideration. In table II we provide the obtained results  together with the corresponding errors that arise from the uncertainties inherited from the input parameters, $ b $ quark mass as well as from  ambiguities of the working intervals of  the auxiliary parameters.  It is worth nothing that, as is seen from table II, there is a large mass splitting ($ \sim 960  $ MeV) between the central values of two opposite parities in spin-$ 5/2 $ channel  compared to the ones of spin-$ 3/2 $ states ($ \sim 30  $ MeV). This can be attributed to the different interpolating currents and internal structures used in these channels. For the spin-$ 3/2 $ states we considered the molecular structure $ \bar{B}^*\Sigma_{b} $, while for the spin-$ 5/2 $ states we used the admixture of the $\bar{B}\Sigma_{b}^{*}$ and $\bar{B}^{*}\Lambda_{b}$ molecular structures with a mixing angle that we fixed later. 
\begin{table}[tbp]
\begin{tabular}{|c|c|c|}
\hline\hline
       $ J^{P} $  &  $m~(\mathrm{GeV})$ & $\lambda~(\mathrm{GeV}^{6})$ \\ \hline\hline
$\frac{3}{2}^{+}$ &  $10.93^{+0.82}_{-0.85}$ & $ (0.22^{+0.04}_{-0.04})\times 10^{-2}$     \\
\hline
$\frac{3}{2}^{-}$ &  $10.96^{+0.84}_{-0.88}$ & $ (0.36^{+0.05}_{-0.05})\times 10^{-2}$     \\
\hline
$\frac{5}{2}^{+}$ &  $11.94^{+0.84}_{-0.82}$ & $ (0.60^{+0.15}_{-0.16})\times 10^{-2}$     \\
\hline
$\frac{5}{2}^{-}$ &  $10.98^{+0.82}_{-0.82}$ & $ (0.19^{+0.04}_{-0.03})\times 10^{-2}$     \\
\hline\hline
\end{tabular}%
\caption{The results of QCD sum rules calculations for the mass and residue of the bottom pentaquark states.}
\label{tab:Values}
\end{table}

We would also like to compare our predictions  with the existing results of other studies on  $ 3/2^- $ and $5/2^+  $ bottom pentaquarks states. In Ref.~\cite{Chen:2015moa} the values for the masses are obtained as $m_{[\bar{B}^*\Sigma_b],3/2^-}=11.55^{+0.23}_{-0.14}$~GeV and $m_{[\bar{B}\Sigma_b^*  \bar{B}^*\Lambda_b],5/2^+}=11.66^{+0.28}_{-0.27}$~GeV. Though our  predictions for  the masses of these states are in  agreements with the results of Ref.~\cite{Chen:2015moa} considering the errors, the central value in our case is considerably low (high) for $J^P=3/2^-$ ($J^P=5/2^+$) state compared to the predictions of Ref.~\cite{Chen:2015moa}. Our results on the residues as well as the masses of the opposite-parity states can be checked via different theoretical approaches. The results of this work on the masses may shed light on  future experimental searches  especially those at LHCb. 

\section{Summary and Outlook}
In this work the masses and residues of the hidden bottom pentaquarks with  quantum numbers $J=3/2$ and $J=5/2$ and both the positive and negative parities have been computed using the QCD sum rule method. We adopted a molecular current of the $\bar{B}^*  $ meson and $  \Sigma_{b}$ baryon to explore the states with $J=3/2$, while a mixed molecular current of $ \bar{B} $ meson and $ \Sigma_{b}^{*} $ baryon with $  \bar{B}^{*}$ meson and $ \Lambda_{b} $ baryon have been used to interpolate the states with $J=5/2$  and both parities. After fixing the auxiliary parameters, namely the continuum threshold and Borel parameter for both the spin-3/2 and spin-5/2 states as well as  the mixing parameter in spin-5/2 channel we found the numerical  values of the masses and residues and   compared the obtained results on the masses with the existing results of other theoretical studies. Although our  predictions for  the masses of the negative parity spin-3/2 and positive parity spin-5/2 states are in nice consistencies with the results of Ref.~\cite{Chen:2015moa} considering the uncertainties, the central value in our case is considerably low (high) for $J^P=3/2^-$ ($J^P=5/2^+$) state compared to the predictions of Ref.~\cite{Chen:2015moa}. Our results on the masses of the opposite parity states as well as the residues can be verified via different theoretical studies. These results  may shed light on the future  experimental searches especially those that are  conducted at LHCb.  

 Our predictions for the masses of the considered states allow us to consider  the decay modes like the $S$-wave $\Upsilon(1S)N$, $\Upsilon(2S)N$, $\Upsilon(1S)N(1440)$, $\Upsilon(1D)N$  and possibly $\bar{B}^*\Sigma_b$ decay channels  for the  spin-3/2 hidden bottom pentaquark states, as well as the  $S$-wave $\Upsilon(1S)\Delta$, $P$-wave $\bar{B}^*\Lambda_b$, $\bar{B}^*\Sigma_b$, $\Upsilon(1S)N$, $\Upsilon(2S)N$, $\Upsilon(1s)N(1440)$, $\psi_{b1}(P)N$, $h_b(1P) N$  and $D$-wave $\Lambda_bB$  channels for the  spin-5/2 decays. Investigation of these decay channels may provide valuable information for the experimental studies and help one to understand the structure of these particles, as well as their interaction mechanisms. We shall use our present results for the masses
and residues of the pentaquarks  in our future studies to analyze such strong decay
channels.

\section*{ACKNOWLEDGEMENTS}

K. A. and Y. S. thank  T\"{U}B\.{I}TAK for partial support provided under the Grant no: 115F183.
The work of H. S. was supported partly by BAP grant 2017/018 of Kocaeli University. The authors would also like to thank S. S. Agaev for his useful discussions.

\label{sec:Num}

\section*{APPENDIX: SPECTRAL DENSITIES}

As examples, in this appendix,  we present the perturbative and nonperturbative parts of the spectral densities corresponding to the structures $g_{\mu\nu}$  and  $  (g_{\mu\rho}g_{\nu\sigma}+g_{\mu\sigma}g_{\nu\rho})/2 $ in terms of the the integrals over the Feynman parameters $x$ and $ y $:

\begin{widetext}
\begin{eqnarray} 
\rho_{\frac{3}{2}}^{\mathrm{m,pert}}(s)&=&\frac{m_b}{5\times 2^{15}\pi ^{8}}\int\limits_{0}^{1} dx\int\limits_{0}^{1-x}dy \frac{\left(6sw-m_b^2r \right)\left(sw-m_b^2r \right) ^{4} }{h^3 t^8} \Theta\left[L \right] , 
 \notag \\
\rho^{m}_{\frac{3}{2},\mathrm{3}}(s)&=&\frac{m_b^2}{2^{9} \pi ^6}\langle
\bar{d}d\rangle\int\limits_{0}^{1} dx\int\limits_{0}^{1-x}dy \frac{\left( sw-m_b^2t(x+y)\right) ^{3}}{h^2t^5}\Theta\left[L \right],  
   \notag \\
\rho^{m}_{\frac{3}{2},\mathrm{4}}(s)&=&\langle\frac{\alpha_{s}}{\pi}G^{2} \rangle \frac{1}{3^{2} \times 2^{15} \pi ^6}\int\limits_{0}^{1} dx\int\limits_{0}^{1-x}dy \frac{\left[sw-m_b^2t(x+y) \right] }{h^3 t^7}\left\lbrace 12 m_b s w y^2(h^2 s x^3+m_b^2 t^2 y)\right. 
 \notag \\
&-& \left. 6m_b y \left( m_b^2 t (x+y)-sw\right)\left[2h^2 sx^3y+m_b^2t^2y^2+hsx \left(34x^4+2y(y-1)^2(16y-9)
\right. \right.\right.
 \notag \\
&+& \left.\left.\left.
x^3(105y-88)+x^2(72-209y+137y^2)+2x(50y^3-102y^2+61y-9) \right)   \right]
\right.
 \notag \\
&+& \left.
m_b\left(sw-m_b^2t(x+y) \right) ^2 \left[6h^2y^2+\left( 68x^4+3y(y-1)^2(17y-12)
\right.\right.\right.
 \notag \\
&+& \left.\left.\left.
x^3(197y-176)+8x^2(18-49y+31y^2)+3x(58y^3-123y^2+77y-12)\right)  \right]   
 \right\rbrace  
\Theta\left[L \right],  \notag \\
\rho^{m}_{\frac{3}{2},\mathrm{5}}(s)&=&\frac{3m_b^2}{2^{10} \pi^6}m_0^2\langle
\bar{d}d\rangle\int\limits_{0}^{1} dx\int\limits_{0}^{1-x}dy\frac{\left( sw-m_b^2t(x+y)\right)^{2} }{ht^4}\Theta\left[L \right],
\notag \\
\rho^{m}_{\frac{3}{2},\mathrm{6}}(s)&=& \frac{m_b}{3^3\times 2^{8}\pi^6}\left(2g_s^{2} \langle\bar{u}u\rangle^{2}+g_s^{2} \langle\bar{d}d\rangle^{2} \right) \int\limits_{0}^{1} dx\int\limits_{0}^{1-x}dy \frac{x\left(m_b^2r- 3sw\right) \left(m_b^2r- sw\right)}{t^5} \Theta\left[L \right]
\notag \\
&+&\frac{m_b}{ 2^{4}\pi^4}\langle\bar{u}u\rangle^{2}\int\limits_{0}^{1} dx\int\limits_{0}^{1-x}dy \frac{x\left(m_b^2r- 3sw\right) \left(m_b^2r- sw\right)}{t^5} \Theta\left[L \right],\nonumber\\
\rho_{\frac{5}{2}} ^{\mathrm{m,pert}}(s)&=&\frac{m_b\left(5 \cos^2\theta-4 \cos \theta \sin\theta+12 \sin^2\theta \right) }{2^{17}\times 3 \times 5^{2}\pi ^{8}}\int\limits_{0}^{1} dx\int\limits_{0}^{1-x}dy  \frac{ x \left(5 x^2+x (y+5z)+ 5 z y\right)}{h^3 t^9}
\nonumber\\
&\times&
 \left(s w-m_b^2 r\right)^4
\left(m_b^2 r-
6 s w\right)\Theta\left[L \right] ,  \notag \\
\rho^{m} _{\frac{5}{2},\mathrm{3}}(s)&=&-\frac{m_b^2\left(\cos^{2}\theta
   (\langle
\overline{d}d\rangle+4 \langle
\overline{u}u\rangle)+4 \cos \theta \sin \theta (\langle
\overline{d}d\rangle-2
   \langle
\overline{u}u\rangle)\right)}{2^{11}\times 3^{2} \times \pi ^6}\int\limits_{0}^{1} dx\int\limits_{0}^{1-x}dy\frac{ \left(3 x^2+x (y+3z)+3yz\right) }{h^2 t^6} \nonumber\\
&\times&\left(m_b^2 r-s w\right)^3\Theta\left[L \right],
   \notag \\
\rho^{m} _{\frac{5}{2},\mathrm{4}}(s)&=&-\langle\frac{\alpha_{s}}{\pi}G^{2} \rangle \frac{m_b}{2^{17}\times 3^{3}\times 5 \pi ^6}\int\limits_{0}^{1} dx\int\limits_{0}^{1-x}dy\frac{x (m_b^2 r-s w) }{h^3t^8}
\Bigg\{4 \cos \theta \sin \theta\Big(4 s^2 w^2 \Big(20 x^6 + 100 z^3 y^3
\nonumber\\
&+&
 4 x^5 (31 y+10z) + 5 x z^2 y^2 (56z + 27 y)
     +40 x^3 y (13 - 33 y + 20 y^2) + x^4 (20 - 504 y + 505 y^2) +5 x^2 y
\nonumber\\ &\times&
      (219 y - 337 y^2 + 154 y^3-36)\Big) +
  m_b^4 t^2 \Big (20 x^8 +
     10 z^2 y^5 (22z + 3 y) + x^7 (40z + 314 y) +
     x^6 (20 - 1004 y )
\nonumber\\ &+&
    1639 y^2 +
     2 x^5 y (475 - 2192 y + 1956 y^2) +
     3 x y^4 (1095 y - 1232 y^2 + 457 y^3-320 )
    + x^2 y^3 ( 6525 y-8537y^{2}
    \nonumber\\ &+&
     3572 y^3-1560) +
     x^3 y^2 (-1120 + 6865 y - 11362 y^2 + 5623 y^3) +
     x^4 y (-300 + 3865 y - 9221 y^2 + 5779 y^3)\Big)
\nonumber\\ &-&
  m_b^2 s x y \Big(100 x^{10} + 10 z^4 y^5 (62z + 9 y) +
     10 x^9 (40y + 93 y) +
     x z^3 y^4 (2880 - 7505 y + 4733 y^2)
\nonumber\\ &+&
     x^8 (600 - 6220 y + 6633 y^2) +
     x^2 z^2 y^3 ( 23645 y -4920- 34196 y^2 +  15489 y^3) +
     x^7 (  11700 y -400
    \nonumber\\ &-&
      29884 y^2 + 18857 y^3) +
     x^3 z^2 y^2 (-3680 + 29025 y - 56766 y^2 + 31998 y^3) +
     2 x^6 (50 - 5540 y
     \nonumber\\ &+&
      26777 y^2 - 39055 y^3 + 17777 y^4) +
     2 x^5 y (2645 - 23834 y + 63097 y^2 - 65643 y^3 + 23735 y^4) \nonumber\\ &+&
     x^4 y (-1020 + 21045 y - 98406 y^2 + 183623 y^3 - 151144 y^4 +
        45902 y^5))\Big)\nonumber
   \end{eqnarray}
\begin{eqnarray} 
       &+& 24 \sin^{2} \theta \Big(m_b^4 t^2 (20 x^8 + x^7 (83z-17) -
    15 z^2 y^4 ( 3 y^2-2)+
    x^3 y (120 - 750 y + 895 y^2 + 256 y^3 - 524 y^4) \nonumber\\&+& x^6 (170 - 398 y + 113 y^2) -
    x^5 (120 - 665 y + 623 y^2 + 36 y^3)+x^2 y^2 (180 - 630 y + 345 y^2 + 541 y^3 - 436 y^4)  \nonumber\\&+&
    x^4 (30 - 470 y + 1080 y^2 - 347 y^3 - 332 y^4) +
    x y^3 (120 - 290 y+20 y^2 + 353 y^3 - 203 y^4)\Big)  \nonumber\\&+&
 4 s^2 w^2 \Big(20 x^6 - 30 z^3 y^2 +
    4 x^5 (22 z-3) + 10 x z^2 y (6 - 11 y + y^2) +
    x^4 (170 - 318 y + 145 y^2)
\nonumber\\
&+&
    5 x^3 (-24 + 86 y - 89 y^2 + 27 y^3) +
    10 x^2 (3 - 26 y + 50 y^2 - 33 y^3 + 6 y^4)\Big)\nonumber\\&-&
 m_b^2 s x y \Big(100 x^{10} + 35 x^9 ( 19 z-1) -
    15 z^4 y^4 ( 8 y + 5 y^2-10) +
    6 x^8 (325- 690 y + 336 y^2)\nonumber\\
&-&
    2 x z^3 y^3 (300 - 740 y + 300 y^2 + 167 y^3) +
    2 x^7 (-1400 + 5225 y - 5704 y^2 + 1837 y^3) \nonumber\\&-&
    x^3 z^2 y ( 5580 y - 11835 y^2 + 6772 y^3 +
       319 y^4-600) -
    x^2 z^2 y^2 ( 4620 y - 6505 y^2 2152 y^3 +
       642 y^4\nonumber\\&-&900 ) +
    x^6 (2200 - 13760 y + 26198 y^2 - 19000 y^3 + 4353 y^4) +
    x^5 (10005 y - 31216 y^2+39533 y^3-900 \nonumber\\&-&   20672 y^4 +
       3250 y^5) +
    2 x^4 (75 - 1910 y + 10145 y^2 - 20991 y^3 + 19413 y^4 -
       7324 y^5 + 592 y^6)\Big)\Big)\nonumber\\&+& 
       5 \cos^2\theta\Big(4 s^2 w^2 \Big(52 x^6 +
     4 z^3 y^2 ( 5 z-13) + 4 x^5 ( 61 z-1) +
     x z^2 y (144-320 y + 107 y^2) +x^4\nonumber\\&\times&
     (412 - 864 y + 449 y^2) -
     4 x^3 (72 - 284 y + 333 y^2 - 121 y^3) +
     x^2 (72 - 660 y + 1419 y^2 - 1129 y^3 + 298 y^4)\Big) \nonumber\\&+&
  m_b^4 t^2 \Big(52 x^8 +
     x^7 ( 270 z+22) - 2 z^2 y^4 (  22 y + 29 y^2) +
     x^6 (412 - 1156 y + 599 y^2-36) \nonumber\\&+&
     2 x^5 ( 893 y - 1186 y^2 + 348 y^3-144) +
     x^2 y^2 (432 - 1824 y + 2133 y^2 - 409 y^3 - 332 y^4) \nonumber\\&+&
     x^3 y (288 - 2024 y + 3521 y^2 - 1658 y^3 - 133 y^4) -
     3 x y^3 (296 y - 235 y^2 -36 y^3 + 71 y^4-96) \nonumber\\&+&
     x^4 (72 - 1188 y + 3365 y^2 - 2677 y^3 + 359 y^4)) -
  m_b^2 s x y (260 x^{10} + 2 x^9 (-880 + 931 y) \nonumber\\&-&
     2 z^4 y^4 (  206 y + 19 y^2-180) +
     5 x^8 (960 - 2236 y + 1233 y^2) +
     x z^3 y^3 ( 4128 y -2941 y^2-1440 \nonumber\\&+&   145 y^3) +
     x^7 (  27420 y - 33356 y^2 + 12589 y^3-6800) +
     x^2 z^2 y^2 (2160 - 12072 y + 20341 y^2-12004y^3 \nonumber\\&+&
        1557 y^4) +
     x^3 z^2 y (1440 - 14128 y + 34209 y^2 - 27606 y^3 +
        5634 y^4) +
     2 x^6 (2650 - 17620 y \nonumber\\&+& 36793 y^2 - 30611 y^3 + 8779 y^4) +
     2 x^5 (  12535 y - 42226 y^2 + 60059 y^3 - 37935 y^4 +
        8647 y^5-1080) \nonumber\\&+&
     x^4 (360 - 9372 y + 52905 y^2 - 120438 y^3 + 129907 y^4 -
        65384 y^5 + 12022 y^6)\Big)\Big)\Bigg\} \Theta\left[L \right] ,  \notag \\\nonumber\
   \end{eqnarray}
\begin{eqnarray} 
\rho^{m} _{\frac{5}{2},\mathrm{5}}(s)&=&\frac{\left( \cos \theta-2\sin \theta\right) m_b^2 m_0^2\left( 6 \sin \theta \langle
\overline{d}d\rangle + \cos \theta (\langle
\overline{d}d\rangle +4 \langle
\overline{u}u\rangle)\right) }{2^{13} \pi^6}\int\limits_{0}^{1} dx\int\limits_{0}^{1-x}dy\frac{\left( s w-m_b^2 t (x+y)\right)^{2} }{h t^5}
\nonumber \\
&\times&\left( 2x^2 +x(3z+1)+2yz\right) 
\Theta\left[L \right],\nonumber\\
\rho ^{m}_{\frac{5}{2},\mathrm{6}}(s)&=&\int\limits_{0}^{1} dx\int\limits_{0}^{1-x}dy\left\lbrace\left( 2g_s^{2} \langle\overline{u}u\rangle^{2}+g_s^{2} \langle\overline{d}d\rangle^{2}\right)  \frac{m_b\left(5 \cos^2 \theta-4 \cos \theta \sin \theta+12 \sin^2 \theta\right) }{2^{11}\times3^{4} \pi^6}\left(m_b^2t(x+y)-3sw \right)\right.
\nonumber\\
      &\times& \left.
\left(m_b^2t(x+y)-shxy \right)\left( 2xyz+x^2(2x+3y-2)\right)
- \frac{m_b\left(\cos \theta-2 \sin \theta \right)}{3\times 2^{8}\pi^4 t^5}\left[ \langle\overline{u}u\rangle^{2}  \left(\cos \theta+6\sin \theta \right) +4\langle\overline{u}u\rangle \langle\overline{d}d\rangle \cos \theta\right]  \right.
\nonumber\\
      &\times& \left.\left(  m_b^2tx(x+y)-3s w x\right)\left(m_b^2t(x+y)-sw \right)   \right\rbrace \Theta\left[L \right],
\end{eqnarray}
where $ \Theta\left[L \right] $ is the usual unit-step function and we have  used the  shorthand notations 
\begin{eqnarray}
z&=&y-1,
\nonumber\\
h&=&x+y-1,
\nonumber\\
t&=&x^2+(x+y) (y-1),
\nonumber\\
r&=&x^3+x^2 (2 y-1)+ y(y-1)(2x+y),
\nonumber\\
w&=&hxy,
\nonumber\\
L&=&\frac{z}{t^2}\left[sw-m_b^2(x+y)t\right] .
\end{eqnarray}
\end{widetext}


\end{document}